# Unidirectional light propagation through two-layer nanostructures based on optical near-field interactions

Makoto Naruse,[1,*] Hirokazu Hori,[2] Satoshi Ishii,[3] Aurélien Drezet,[4] Serge Huant,[4] Morihisa Hoga,[5] Yasuyuki Ohyagi,[5] Tsutomu Matsumoto,[6] Naoya Tate,[7] and Motoichi Ohtsu[7]

1 Photonic Network Research Institute, National Institute of Information and Communications Technology, 4-2-1 Nukui-kita, Koganei, Tokyo 184-8795, Japan

2 University of Yamanashi, Takeda, Kofu, Yamanashi 400-8511, Japan

3 Advanced ICT Research Institute, National Institute of Information and Communications Technology, 588-2, Iwaoka, Nishi-ku, Kobe, Hyogo 651-2492, Japan

4 Institut Néel, CNRS and Université Joseph Fourier, 25 rue des Martyrs BP 166, 38042 Grenoble Cedex 9, France

5 Dai Nippon Printing Co. Ltd., 250-1 Wakashiba, Kashiwa, Chiba 277-0871, Japan

6 Yokohama National University, Hodogaya, Yokohama, Kanagawa 240-8501, Japan

7 Department of Electrical Engineering and Information Systems, Graduate School of Engineering, The University of Tokyo, 2-11-16 Yayoi, Bunkyo-ku, Tokyo 113-8656, Japan

*Corresponding author: Electronic mail: naruse@nict.go.jp

**Abstract:** We theoretically demonstrate direction-dependent polarization conversion efficiency, yielding unidirectional light transmission, through a two-layer nanostructure by using the angular spectrum representation of optical near-fields. The theory provides results that are consistent with electromagnetic numerical simulations. This study reveals that optical near-field interactions among nanostructured matter can provide unique optical properties, such as the unidirectionality observed here, and offers fundamental guiding principles for understanding and engineering nanostructures for realizing novel functionalities.

# 1. INTRODUCTION

Unidirectional light propagation has been intensively studied because of its crucial importance in practical optical systems, such as to achieve one-way signal transfer or avoid back-reflections. Nonreciprocal light propagation [1,2] has been achieved in magneto-optical materials. The Faraday effect is the most well-known example [3], and this effect is used in a variety of device architectures, such as waveguides with different propagation constants in the forward and backward directions [4], different propagation losses [5], and polarization conversion [6], have been demonstrated. By using isotropic materials, Lockyear *et al.* demonstrated a one-way diffraction grating that exhibits diffracted beams from one surface only [7], and unidirectional light transmission has also been shown [8,9]. Asymmetric transmission of linearly or circularly polarized light by optical metamaterials has also been demonstrated [10, 11,12].

Regarding isotropic-material-based nanostructures that exhibit unidirectional/asymmetric light transmission [7-9,12], a common feature is that they contain subwavelength-scale three-dimensional architectures, typically consisting of two-layer structures. Interactions among nanostructured two-layer systems yield interesting optical properties. Nevertheless, the physical reasoning and formalism have been limited to the notion of far-field optics [12,13]. The coupling of surface plasmon polaritons has been postulated as the elemental physical process [9]; however, a detailed formalism concerning near-field processes and unidirectional transmission has not been investigated yet.

In this paper, we present a rigorous theoretical foundation for characterizing unidirectional signal transfer in two-layer nanostructured matter based on the angular spectrum representation of optical near-fields. The evanescent wave is the most remarkable manifestation of optical near-fields that propagate parallel to the boundary surface and exhibit an exponential decay in the direction normal to the surface [14-16]. In the near-field regime, the angular spectrum of the scattered fields involves evanescent waves with wavevectors (or momentum) along the surface much larger than that of optical wave in free space [17]. In addition, with respect to an arbitrary assumed planar boundary, the angular spectrum representation of optical near-fields is a very useful basis. Therefore, the angular spectrum theory of optical near-fields has been successful in revealing multipole radiation near a planar boundary [16], localized hierarchical optical interactions on the subwavelength scale [18], and asymmetric polarization conversion [19].

What we particularly address in this paper is to theoretically and numerically demonstrate that the polarization conversion efficiency from *x*-polarized input light to *y*-polarized output light in the *forward* direction differs from the polarization conversion efficiency from *y*-polarization input to *x*-polarization output in the *backward* direction through a two-layer planar nanostructure. By considering the polarization conversion in the forward and backward directions based on the angular spectrum theory of optical near-fields, the unidirectional transmission can clearly be grasped. Also, the ability of the angular-spectrum-based theory to explicitly deal with multiple multipoles, including dipoles, that are located at arbitrary positions is fully utilized in characterizing the relevance between geometrical features of nanostructures and the resultant direction-dependent polarization properties.

Before proceeding, it is instructive to consider the intuitive physical meaning of the angular spectrum with respect to the polarization conversion problem mentioned above. Consider the dispersion relation of photons,

$$\left(\frac{\omega}{c}\right)^2 - k_\parallel^2 - k_\perp^2 = 0, \tag{1}$$

where $\omega$ is the temporal angular frequency, and $k_\parallel$ and $k_\perp$ are wavenumbers parallel and perpendicular to an assumed planar boundary [20]. When the wavenumber parallel to the boundary is larger than that of free space, that is, $k_\parallel^2 > k^2 = k_\parallel^2 + k_\perp^2$, then $k_\perp^2 < 0$ should hold, meaning that the wavenumber perpendicular to the boundary is an imaginary number, which corresponds to evanescent waves. The angular spectrum representation is a momentum-space expansion of arbitrary scattered fields with respect to an assumed planar boundary in a series of plane waves that include complex wavenumbers. The exchange or transfer of momentum (or wavenumber) parallel to the boundary ($k_\parallel$), which is much larger than that in free space regarding near-field components, is explicitly dealt with by the angular spectrum formulation. It should also be noted that the large momentum originates from polarizations in the nanostructured matter induced by incident light in free-space, and it is also correlated with scattered light or radiation from the nanostructure. In this paper, the problem of polarization conversion in two-layer nanostructured matter is analyzed as a "filtering" of evanescent components, or transfer of momentum, with respect to an assumed planar boundary. The asymmetric architecture of the nanostructure is explicitly incorporated into the angular spectrum-based formulation, and the direction-dependent polarization conversion and irreversibility of light transmission involving optical near-field processes is clearly described.

This paper is organized as follows. In Section 2, the theoretical elements and an outline of the method of dealing with unidirectionality are presented. Section 3 describes a concrete example of a two-layer nanostructured system in which the above-mentioned direction-dependent polarization conversion takes place. Section 4 presents a theoretical analysis of the polarization conversion shown in Section 3. Section 5 concludes the paper.

## 2. THE ANGULAR SPECTRUM REPRESENTATION OF OPTICAL NEAR-FIELDS: THE BASIS OF DIRECTION-DEPENDENT POLARIZATION CONVERSION

In order to deal with light scattering based on subwavelength-scale interactions involving exchange of momentum, it is convenient to expand the vector multipole field into vector plane waves with respect to an arbitrary assumed planar boundary by means of the angular spectrum representation [16]. The state of vector plane waves propagating in an arbitrary direction is characterized by wavenumber $K$, the directional angles of the wavevector $\alpha$ and $\beta$, and the polarization state $\mu$ (for a transverse electric wave, $\mu =$ TE, and for a transverse magnetic wave, $\mu =$ TM) with respect to the wavevector. The total angular momentum is denoted by $J$, and its $z$-projection (magnetic quantum number) is denoted by $m$. The vector plane wave is described in the form $\varepsilon(s^{(+)},\mu)\exp(iKs^{(+)}\cdot r)$ ($\mu =$ TE, TM) where

$$
\begin{aligned}
&s^{(+)} = (s_x, s_y, s_z) = (\sin\alpha\cos\beta, \sin\alpha\sin\beta, \cos\alpha) \\
&\varepsilon(s^{(+)}, \text{TM}) = (\cos\alpha\cos\beta, \cos\alpha\sin\beta, -\sin\alpha) \\
&\varepsilon(s^{(+)}, \text{TE}) = (-\sin\beta, \cos\beta, 0)
\end{aligned} \quad (2)
$$

which are schematically illustrated in Fig. 1(a). Then, the vector spherical waves are expanded into a vector plane wave:

$$
A_{K,J,m}(r) = \frac{1}{2\pi} \sum_{\mu=\text{TE}}^{\text{TM}} \int_{C+} \int_{-\pi}^{\pi} d\Omega_S(\mu,\alpha,\beta|\lambda,J,m)\varepsilon(s^{(+)},\mu)\exp(iKs^{(+)}\cdot r), \quad (3)
$$

where $d\Omega_S = \sin\alpha\, d\alpha\, d\beta$ [15,16]. The expansion coefficient $(\mu,\alpha,\beta|\lambda,J,m)$ is given in Ref. [16]. This corresponds to the angular spectrum representation of a multipole. What is remarkable is that whereas the angle $\beta$ is real ($-\pi \leq \beta < \pi$), $\alpha$ takes a complex value, leading to an explicit and intuitive expression for the optical near-field. In the particular case of this paper, $\alpha$ follows a contour shown by $C_+$ in Fig. 1(b). Regarding the contour $C_+$, the region $0 \leq \text{Re}(\alpha) < \pi/2$ corresponds to a homogeneous mode, and the region $\text{Im}(\alpha) \leq 0$ corresponds to an evanescent mode. It is useful to introduce $s_\parallel = \sin\alpha$, which allows us to write a parameter $s_z$ of the wavevector $\boldsymbol{s}^{(+)}$ as

$$s_z = \begin{cases} \sqrt{1-s_\parallel^2} & \text{for } 0 \leq s_\parallel < 1 \\ i\sqrt{s_\parallel^2-1} & \text{for } 1 \leq s_\parallel < +\infty, \end{cases} \tag{4}$$

meaning that pure imaginary values of $s_z$ describe an evanescent wave propagating parallel to the boundary plane and showing an exponential decay with increasing distance from the boundary. A remarkable benefit is that the value of $s_\parallel$ specifies the property of a plane wave as a homogeneous wave ($0 \leq s_\parallel < 1$) or an evanescent wave ($1 \leq s_\parallel < +\infty$). In the case of an electric dipole $\boldsymbol{d}$ as the source, the electric field is given by [17]

$$\boldsymbol{E}(\boldsymbol{r}) = \left(\frac{iK^3}{8\pi^2\varepsilon_0}\right) \sum_{\mu=TE}^{TM} \int_0^{2\pi} d\beta \int_0^\infty ds_\parallel \frac{s_\parallel}{s_z} \left[\boldsymbol{\varepsilon}(\boldsymbol{s}^{(+)},\mu)\cdot\boldsymbol{d}\right] \boldsymbol{\varepsilon}(\boldsymbol{s}^{(+)},\mu) \exp(iK\boldsymbol{s}^{(+)}\cdot\boldsymbol{r}). \tag{5}$$

We investigate the angular spectrum derived from eq. (5) in the case of multiple dipoles arranged on a subwavelength scale in Section 4. In this section, we outline the essential idea of representing the direction-dependent polarization, which is followed by electromagnetic simulations in Section 3, and concrete theoretical applications in Section 4. Since we consider optical interactions in the subwavelength regime where evanescent components are dominant and homogeneous ones are negligible, we characterize the angular spectra in the regime $1 \leq s_\parallel < +\infty$. We refer to $s_\parallel$ as "spatial frequency" hereafter.

(1) Suppose that $x$-polarized input light in the forward direction induces an electric dipole $\boldsymbol{d}^{(1)}$.

We can derive an angular spectrum representation of the radiation originating from the dipole $\boldsymbol{d}^{(1)}$ at a plane at a distance $Z_M$ from the source (Fig. 1(c)). The forward-direction input light polarization may excite other dipoles at other positions, such as $d^{(1')}$ illustrated in Fig. 1(c), whose angular spectra are also simultaneously considered at the same boundary. In this manner, the angular spectrum triggered by the forward-direction $x$-polarized input light is obtained. Fig. 1(d, i) shows a schematic illustration of such an angular spectrum as a function of spatial frequency $s_\parallel$. One remark to make here is that Fig. 1(c) is a "conceptual" schematic; the dipoles $d^{(1)}$ and $d^{(1')}$ are located on the same plane, and thus the abscissa represented by the dashed line means the spatial position, whereas the decaying solid curves and the array of vertical lines shown in the middle *conceptually* indicate the exponentially decaying nature and high spatial frequency of evanescent components of the electric fields, respectively.

(2) At the same time, we consider that "ideal" $y$-polarized output light is related with the radiation from an electric dipole $d^{(2)}$. The angular spectrum originating from the dipole $d^{(2)}$ can also be considered at a virtual boundary plane corresponding to the one introduced at Step (1). The "ideal" output light may also involve other dipole(s) (not shown in Fig. 1(c)) whose angular spectra are considered as well. We call the corresponding angular spectrum a "filter" angular spectrum (Fig. 1(d,ii)).

(3) In characterizing the amount of $y$-polarized output light in the forward direction originating from the $x$-polarized input light, we consider that a limited portion of the angular spectrum triggered by the forward-direction input light is "filtered" by the angular spectrum specified in Step (2) (Fig. 1(d,iii)). Suppose that the polarization conversion from the $x$-polarization to the $y$-polarization takes place perfectly. Then the entire angular spectrum given in Step (1) should correspond to the angular spectrum given in Step (2). As described in the introduction, the angular spectrum corresponds to momentum parallel to the assumed boundary, and thus the perfect conversion efficiency means a perfect exchange of momentum for all wavenumbers, whereas less-than-perfect conversion indicates that the momentum exchange is suppressed at certain wavenumbers.

(4) Next, we consider the backward direction. The *y*-polarized input light of the backward direction excites an electric dipole $\boldsymbol{d}^{(2)}$. Note that this dipole $\boldsymbol{d}^{(2)}$ is equivalent to the dipole $\boldsymbol{d}^{(2)}$ considered in Step (2). In this manner, the angular spectrum triggered by the backward-direction input light is obtained (Fig. 1(d,iv)).

(5) In the backward direction, by following exactly the same formalism regarding the forward direction shown in Step (2), the "filter" is specified by $\boldsymbol{d}^{(1)}$ (and $\boldsymbol{d}^{(1')}$) that yields *x*-polarized output light in the backward direction (Fig. 1(d,v)).

(6) The output light polarization in the backward direction is characterized by the angular spectrum such that a portion of the input angular spectrum, obtained in Step (4), is chosen by the filter specified by the angular spectrum given in Step (5) (Fig. 1(d, vi)).

The "unidirectionality" means that the angular spectra of the forward and backward output light could differ from each other, which is denoted by the letter (U) in Fig. 1(d). Moreover, if the input angular spectrum of the forward direction differs from that of the output angular spectrum of the backward direction, the system could be regarded as having *irreversibility* or *nonreciprocity*, denoted by the letter (N) in Fig. 1(d) in the sense that the forward-direction input light polarization is not retrievable when it passes through the system in the forward direction and comes back in the backward direction via optical near-field interactions.

The notion of "nonreciprocity", however, has been referred to as a phenomenon that breaks Lorentz reciprocity in the literature [21]; the reciprocity theorem holds provided that the electrical permittivity and magnetic permeability are symmetric. We should note that our study, at this stage, has not yet addressed the relation between optical near-field processes and the Lorentz reciprocity; this is an important and interesting future issue to be examined in light–matter interactions on the nanometer-scale where the applicability of the notion of macroscale concepts, such as permittivity and permeability, should be investigated in greater depth. Carminati *et al.* studied reciprocity of evanescent electromagnetic waves where generalized reflection and transmission coefficients of vector wave fields containing evanescent components were investigated [22]. Whereas the scatter under study in Ref. [22] was considered as an united system, our study deals with two separated systems divided by a virtual boundary plane interacting via optical near-fields where we introduce the notion of exchange of momentum. Meanwhile, our study deals with limited cases of linearly polarized light with respects to the layout of nanostructures whereas Carminati *et al.* dealt with arbitrarily polarizations. For these reasons, we use the term "unidirectionality", not "nonreciprocity", throughout the rest of this article to remove any ambiguity. Nevertheless, we emphasize that this paper should provide a theoretical foundation for investigating the polarization of light from the viewpoint of optical near-field processes.

## 3. ELECTROMAGNETIC SIMULATIONS

This section proposes a particular two-layer nanostructure consisting of gold (Au) embedded in $SiO_2$ substrates (Fig. 2(a)). Considering realistic experimental two-layer devices consisting of gold nanostructures embedded in substrates, such as the one demonstrated by Tate *et al.* in Ref. [23], we choose a surrounding environment consisting of $SiO_2$. The nanostructure in the first layer, which is the entrance for the forward-direction input light, consists of gratings parallel to the *y*-axis with a horizontal interval of $G_X$ (Fig. 2(b)). The second-layer nanostructure is composed of an array of pairs of L-shaped structures. The horizontal (*x*) and vertical (*y*) lengths of each L-shape are indicated by *L*. The two L-shaped structures in a pair face each other with a relative rotation of 180 degrees and are horizontally separated by a distance $G_L$. These L-shaped pairs are arrayed with an interval of $G_X$ horizontally, which is the same with the inter-grating gap of the first layer, and with an interval of $G_Y$ vertically. The height and width of all gold structures are the same, as respectively indicated by *h* and *w*. The inter-layer gap distance between the first and the second layers is given by *G*.

We calculate optical properties in both the near-field and far-field based on a finite-difference time-domain (FDTD) method [24], using the *Poynting for Optics* software, a product of Fujitsu, Japan. The permittivity of gold is expressed by the Drude model, in which the refractive index and extinction ratio are 0.16 and 3.8, respectively, at a wavelength of 688 nm [25]. The SiO$_2$ substrate is as an isotropic dielectric material with a refractive index of 1.457. The light source is placed 500 nm away from the surface of the first layer with respect to the forward direction, whereas it is placed 500 nm away from the surface of the second layer with respect to the backward direction. We assume periodic boundary conditions at the edges in the *x*- and *y*-directions and perfectly absorbing boundaries in the *z*-direction.

In the representative structures, the parameters of the structures are given the following values. Elemental-structure-related specifications: *w* (width), 80 nm; *h* (thickness), 80 nm; *L*, 200 nm; and $G_L$, 180 nm. Layout-related specifications: $G_X$, 800 nm; $G_Y$, 600 nm; and *G* (inter-layer gap distance), 200 nm.

The far-field optical response is calculated at a plane 2.25 μm away from the surface of the structure opposite to the light source for both the forward and backward directions. We assume an input optical pulse with a differential Gaussian form whose duration is 0.9 fs that covers the wavelength regime studied in this paper. The transmission efficiency is given by calculating the Fourier transform of the electric field at the far-field output plane divided by the Fourier transform of the electric field at the light source. We assume that $SiO_2$ occupies the entire computational area in the calculation; this allows us to characterize the essential optical processes while concerning the environmental effects of $SiO_2$. Another possibility is to assume a vacuum environment, which actually yields similar characteristics except that the wavelength regime where unidirectionality occurs (shown below) is shifted to a shorter wavelength range.

We are interested in the polarization conversion efficiency from *x*-polarized (or *y*-polarized) input light to *y*-polarized (or *x*-polarized) output light for both the forward and backward directions. The "forward" direction is assigned to the case where the input light enters the two-layer device from the first layer. The polarization conversion efficiency from *x*-polarized input light to *y*-polarized output for the forward direction is given by

$$T_{X \to Y}^{(F)}(\omega) = \left| \frac{\hat{E}_{y,output}(\omega)}{\hat{E}_{x,input}(\omega)} \right|, \tag{6}$$

where $\hat{E}_{y,output}(\omega)$ denotes the Fourier transform of the *y*-component of the electric field evaluated at the output plane, and $\hat{E}_{x,input}(\omega)$ shows the Fourier transform of the *x*-component of the input electric field. Similarly, the polarization conversion efficiency from *y*-polarization to *x*-polarization for the forward direction is $T_{Y \to X}^{(F)}$. Likewise, the polarization conversion efficiencies for the backward direction are $T_{X \to Y}^{(B)}$ and $T_{Y \to X}^{(B)}$.

The solid, dotted, dot-dashed, and dashed curves in Fig. 2(c) respectively represent the calculated $T_{X \to Y}^{(F)}$, $T_{Y \to X}^{(F)}$, $T_{X \to Y}^{(B)}$, and $T_{Y \to X}^{(B)}$. Based on the optical reciprocity theorem, the values of $T_{X \to Y}^{(F)}$ and $T_{Y \to X}^{(B)}$ should be the same, and the values of $T_{Y \to X}^{(F)}$ and $T_{X \to Y}^{(B)}$ should be the same; we can observe such a characteristic in the wavelength region longer than about 1200 nm. However, especially in the wavelength region between 900 nm and 1200 nm, the differences of the polarization conversion efficiencies, given by $\left|T_{X \to Y}^{(F)} - T_{Y \to X}^{(B)}\right|$ and $\left|T_{Y \to X}^{(F)} - T_{X \to Y}^{(B)}\right|$, emerge as shown by solid and dashed curves respectively in Fig. 2(d); namely, direction-dependent polarization conversion efficiencies result. If we associate $T_{X \to Y}^{(F)}$ with the forward transmission and $T_{Y \to X}^{(B)}$ with the backward transmission, Fig. 2(d) demonstrates unidirectional light transmission. Regarding the metric $\left|T_{X \to Y}^{(F)} - T_{Y \to X}^{(B)}\right|$, if two polarizers are employed in the system in a crossed-Nicol manner to extract *x*-polarized far-field light on the first layer-side and to extract *y*-polarized far-field light on the second layer-side, the metric represents a performance measure of unidirectionality even for unpolarized light; this should help to give an intuitive understanding of the metric defined above as the unidirectionality in this study.

Before moving to a theoretical investigation based on the angular spectrum, here we note some relevant features. Figure 3(a) characterizes the direction-dependent polarization conversion efficiency ( $\left|T_{X \to Y}^{(F)} - T_{Y \to X}^{(B)}\right|$ ) as a function of the inter-layer gap distance (*G*). Specifically, the average value of $\left|T_{X \to Y}^{(F)} - T_{Y \to X}^{(B)}\right|$ in the spectral region from 900 nm to 1200 nm is quantified, which is called the "average unidirectionality". It should be noted that both too-small and too-large gap distances deteriorate the average unidirectionality; that is, the direction-dependent polarization conversion disappears. The maximum average unidirectionality is obtained around a gap distance of 200 to 300 nm. Such a feature is one of the important concerns in the theoretical treatment to be made.

While keeping the gap distance at 200 nm, Fig. 3(b) characterizes the unidirectionality, given by $\left|T_{X \to Y}^{(F)} - T_{Y \to X}^{(B)}\right|$, as a function of the "layout" regarding the pairs of L-shaped structures

located in the second layer. The solid, dashed, and dotted curves correspond respectively to the specifications [$G_X$: 800, $G_Y$: 600], [$G_X$: 700, $G_Y$: 500], and [$G_X$: 600, $G_Y$: 400], while maintaining other parameter specifications constant (the measurement unit of these values is nanometers). As $G_X$ and $G_Y$ decrease, the waveband that exhibits unidirectionality is blue-shifted. Furthermore, the unidirectionality disappears when the layout has a symmetric configuration, as shown by the dash-dot curve in Fig. 3(b), where $G_X$ and $G_Y$ are both 600 nm.

Figure 4 shows the cross-sectional electric field intensity distributions for the cases when a considerable level of unidirectionality appears. The factors $G_X$ and $G_Y$ are 800 nm and 600 nm, respectively, for the "asymmetric layout", whereas both $G_X$ and $G_Y$ are 600 nm for the "symmetric layout". In order to obtain field distributions, we radiate continuous-wave (CW) input light with a wavelength of 1127 nm, and the refractive index and the extinction ratio of the gold structures are 0.312 and 7.93, respectively [25]. The near-field intensity of the $y$-polarized field ($|E_y|^2$) (Fig. 4(a,i)) and the $x$-polarized field ($|E_x|^2$) (Fig. 4(a,ii)) are evaluated by irradiating forward-direction $x$-polarized input light and backward-direction $y$-polarized input light, respectively. An evident difference between forward and backward directions is observed for the asymmetric layout, whereas no evident interactions are observed in the case of the symmetric layout especially regarding the "forward" direction (Fig. 4(b,i)).

## 4. THEORY OF UNIDIRECTIONAL LIGHT PROPAGATION VIA OPTICAL NEAR-FIELDS BASED ON ANGULAR SPECTRUM REPRESENTATION

Based on the theoretical approach outlined in Section 2, we characterize the direction-dependent polarization conversion, or unidirectionality, by using the angular spectrum representation of optical near-fields. The two-layer system is schematically shown in Fig. 5(a), in which a "*virtual intermediate layer*" is assumed in the middle of the two layers.

With *x*-polarized forward-direction input light, the electron charges in each of the grating structures in the first layer structure are concentrated at the left and right-hand edges with different signs thanks to the coupling between input light and electrons in the metal. We model such a situation by an array of induced oscillating electrical dipoles appearing at the left and right edges of the gratings (Fig. 5(b)). The dipoles are oriented parallel to the *x*-axis, and a pair of dipole located at both edges have a phase difference of $\pi$.

On the other hand, with *y*-polarized backward-direction input light, the electron charges are concentrated at the corners of the L-shaped elemental structures, whereby the signs of the induced electron changes at the corners of a single L-shape structure are opposite from each other. Similarly to the above-mentioned forward-direction input light modeling, we consider that an array of oscillating electrical dipoles is induced at the positions indicated by the arrows shown in Fig. 5(c). Referring to the physical dimensions in the electromagnetic simulations in Section 3, we specify the relative distances between the dipoles as shown in Fig. 5(c); note that the dimensions are given in units of wavelength. Also, in order to equate the power of the total input light for both forward and backward directions, the numbers of induced electrical dipoles are assumed to be the same between the forward *x*-polarized input light and the backward *y*-polarized input light, and the amplitudes of all induced dipoles are also assumed to be the same.

In order to quantify the angular spectra triggered by these arrays of dipoles, we introduce the following model. Suppose that the dipole $\boldsymbol{d}^{(k)}$ is oriented at an angle $\theta^{(k)}$ with respect to the *z* axis and at an angle $\phi^{(k)}$ in the *xy* plane, that is, $\boldsymbol{d}^{(k)} = d^{(k)}(\sin\theta^{(k)}\cos\phi^{(k)}, \sin\theta^{(k)}\sin\phi^{(k)}, \cos\theta^{(k)})$, as schematically shown in Fig. 5(d). Suppose also that we observe the radiation from $\boldsymbol{d}^{(k)}$ at a position displaced from the dipole by $\boldsymbol{R}^{(k)} = (r_\parallel^{(k)} \cos\varphi^{(k)}, r_\parallel^{(k)} \sin\varphi^{(k)}, Z_M)$. In such a case, based on eq. (5), the angular spectrum representation of the $z$-component of the electric field at the position $\boldsymbol{R}$ in the assumed boundary plane for evanescent waves (namely, $1 \leq s_\parallel < +\infty$) is given by [18]:

$$E_z(\boldsymbol{R}) = \left(\frac{iK^3}{4\pi\varepsilon_0}\right) \sum_k \int_1^\infty ds_\parallel \frac{s_\parallel}{s_z} f_z^{(k)}(s_\parallel, \boldsymbol{d}^{(k)}, \boldsymbol{R}^{(k)}), \tag{7}$$

where

$$\begin{aligned} f_z^{(k)}(s_\parallel, \boldsymbol{d}^{(k)}, \boldsymbol{R}^{(k)}) &= ds_\parallel \sqrt{s_\parallel^2 - 1} \sin\theta^{(k)} \cos(\phi^{(k)} - \varphi^{(k)}) J_1\left(Kr_\parallel^{(k)} s_\parallel\right) \exp\left(-KZ_M \sqrt{s_\parallel^2 - 1}\right) \\ &+ ds_\parallel^2 \cos\theta^{(k)} J_0\left(Kr_\parallel^{(k)} s_\parallel\right) \exp\left(-KZ_M \sqrt{s_\parallel^2 - 1}\right). \end{aligned} \tag{8}$$

Here, $J_n(x)$ represents a Bessel function of the first kind, where $n$ is an integer, and the term $\sum_k f_z^{(k)}(s_\parallel, \boldsymbol{d}^{(k)}, \boldsymbol{R}^{(k)})$ corresponds to the angular spectrum of the electric field originating from an array of dipoles $\boldsymbol{d}^{(k)}$.

Here, we focus on the point $P$ at the virtual intermediate layer whose horizontal and vertical position lies directly above the center of the pair of L-shaped elements in the second layer structure (Fig. 5(a)). Furthermore, in order to take account of the layout dependencies of the second layer structure, the angular spectrum concerns dipoles located on the left, right, upper, and lower neighboring areas with respect to the point $P$.

Assuming that $Z_M$ is $\lambda/10$, the angular spectrum at the virtual boundary plane originating from an array of dipoles induced in the first layer by the $x$-polarized forward input light, which is denoted by $f_{X_{IN} \to M}^{(F)}$, is given by eq. (8) and is shown by the solid curve in Fig. 6(i, $Z_M = \lambda/10$). Similarly, an array of dipoles induced in the second layer by the $y$-polarized backward input light gives the angular spectrum denoted by $f^{(B)}_{Y_{IN} \to M}$, as shown by the dashed curve in Fig. 6(i, $Z_M = \lambda/10$).

As outlined in Section 2, a portion of the angular spectrum induced by the forward-direction *x*-polarized input light polarization can contribute to the *y*-polarized output light from the second layer. Ideally, the perfect *y*-polarized forward-direction output light is equivalent to a situation where *y*-polarized backward-direction input light induces an array of dipoles in the second layer. Therefore, we consider that the "filter" that chooses portions that constitute *y*-polarized output light from the angular spectrum $f^{(F)}_{X_{IN} \to M}$ is based on $f^{(B)}_{Y_{IN} \to M}$. Specifically, we normalize $f^{(B)}_{Y_{IN} \to M}$ by either the maximum or minimum value of $f^{(B)}_{Y_{IN} \to M}$ so that the resultant "filter", which is now denoted by $\hat{f}^{(B)}_{Y_{IN} \to M}$, is in the range between [+1, -1] (Fig. 6(ii, $Z_M = \lambda/10$)). As a result, the angular spectrum for the *y*-polarized output light in the forward direction is given by

$$f^{(F)}_{X_{IN} \to Y_{OUT}} = f^{(F)}_{X_{IN} \to M} \times \hat{f}^{(B)}_{Y_{IN} \to M} \qquad (9)$$

where the multiplication is applied for each spatial frequency ($s_\parallel$), and is shown by the solid curve in Fig. 6(iii, $Z_M = \lambda/10$). Following the same approach, the angular spectrum for the *x*-polarized output light in the backward direction is given by

$$f^{(B)}_{Y_{IN} \to X_{OUT}} = f^{(B)}_{Y_{IN} \to M} \times \hat{f}^{(F)}_{X_{IN} \to M}, \qquad (10)$$

where $\hat{f}^{(F)}_{X_{IN} \to M}$ and $f^{(B)}_{Y_{IN} \to X_{OUT}}$ are respectively obtained as the dashed curves in Fig. 6(ii, $Z_M = \lambda/10$) and Fig. 6(iii, $Z_M = \lambda/10$).

It should be noted that the resultant angular spectra $f^{(F)}_{X_{IN} \to Y_{OUT}}$ and $f^{(B)}_{Y_{IN} \to X_{OUT}}$ exhibit different curves, meaning that the polarization conversion efficiency is direction-dependent, which is consistent with the numerical demonstrations shown in Fig. 2(c).

Furthermore, when $Z_M$ is assumed to have either a smaller value ($\lambda/20$) or a larger one ($\lambda/4$), the angular spectra are respectively derived as shown in the row $Z_M = \lambda/20$ and $Z_M = \lambda/4$ in Fig. 6. The differences between the resultant angular spectra of the forward- and backward-directions are smaller than the former case $Z_M = \lambda/10$, which are quantified by calculating $f^{(F)}_{X_{IN} \to Y_{OUT}} / f^{(B)}_{Y_{IN} \to X_{OUT}}$ at each spatial frequency, and we refer to this as the *figure-of-merit (FoM) of unidirectionality*. This FoM can be defined with respect to spatial frequencies for which the denominator is non-zero; the FoM results in a constant value at valid spatial frequencies. This FoM of unidirectionality depends on the inter-layer distance ($2 \times Z_M$), as demonstrated in Fig. 7(a), indicating that the unidirectionality is maximized at distances which are neither too small nor too large; this is consistent with the electromagnetic simulations shown in Fig. 3(a).

Such an attribute is physically reasonable by considering the explicit representation of optical near-fields by the angular spectrum. The angular spectrum representation of optical near-fields, given by eq. (7), indicates that the penetration depth of evanescent waves is very small for larger spatial frequencies ($s_\parallel$), whereas it is large for smaller spatial frequencies; such an attribute is clearly indicated in Fig. 6(i), where the angular spectra can exist at larger spatial frequencies as $Z_M$ decreases. Also, the resultant hierarchical attributes of optical near-fields [18] suggest that near-field effects originating from spatial fine structures cannot be transferred to a distant plane; therefore it is reasonable that the angular spectra of the forward- and backward-directions ($f^{(F)}_{X_{IN} \to Y_{OUT}}$ and $f^{(B)}_{Y_{IN} \to X_{OUT}}$) follow similar traces with larger $Z_M$ ($\lambda/4$). In other words, the near-field effects are negligible as $Z_M$ increases. On the other hand, too-small a value of $Z_M$ indicates that coarse-scale spatial attributes, that is, the asymmetric layout of the dipole arrangements specified by $G_X$ and $G_Y$, are difficult to be delivered through near-field interactions, resulting in the relatively similar angular spectra, as in the case of $Z_M = \lambda/20$.

In order to discuss the layout-dependency, the dotted and dot-dashed curves in Fig. 7(b) respectively indicate the angular spectrum of the forward ($f^{(F)}_{X_{IN} \to Y_{OUT}}$) and backward directions ($f^{(B)}_{Y_{IN} \to X_{OUT}}$) when the second layer L-shapes are arranged in a symmetric manner while $Z_M$ is given by $\lambda/20$. Specifically, both $G_X$ and $G_Y$ (specified in Fig. 5(b) and 5(c)) are $\lambda/2$ for the symmetric architecture. The two angular spectra follow nearly the same traces, meaning that the symmetric layout of the second layer structure causes the unidirectionality to diminish, which is indeed consistent with the simulation results demonstrated in Fig. 3(b). The solid and dashed curves in Fig. 6(b) are respectively the forward and backward angular spectra when the L-shapes are arranged asymmetrically, which behave differently from each other. (These curves are the same as the ones shown in Fig. 6(iii, $Z_M=\lambda/20$)).

Finally, we make a few remarks about the results of the present study. First is a comment regarding the fact that the unidirectionality appears in a particular wavelength range (900–1200 nm), but not at other wavelengths. This is based on the resonance of the input light and the nanostructured matter placed at the entrance side of the input light; in the case of gold embedded in $SiO_2$, the unidirectionality appears in the presented range. Second is a comment about the polarization conversion efficiencies ($T^{(F)}_{X \to Y}, T^{(F)}_{Y \to X}, T^{(B)}_{X \to Y}$, and $T^{(B)}_{Y \to X}$), which take small values in general lower than 10 %. Clarifying the upper bound of the achievable polarization conversion efficiencies and unidirectionality will be an interesting topic of a future study. Also, similarly to the first remark, material-dependence may be present, which could be exploited to increase the unidirectionality. Third is a comment regarding applications of this study. As repeatedly noted in the paper, we consider that the significance of the angular-spectrum-based approach shown here is that we can explicitly grasp the unidirectionality based on the notion of transfer of momentum via optical near-fields. In this context, the proposed principle has advantages in terms of its general-purpose properties or utility. Also, based on the results and/or theoretical elements shown in this paper, we may be able to seek the origin of unidirectionality at a more fundamental level. Meanwhile, as is also mentioned in the literature [1-2,4-13], the potential industrial applications of unidirectional optical devices are vast, including optical isolators, one-way mirrors, etc.

## 5. SUMMARY

In summary, we demonstrate a theoretical foundation for direction-dependent polarization conversion efficiency, yielding unidirectional light transmission, in two-layer isotropic nanostructured matter based on the angular spectrum representation of optical near-fields. The inter-layer distance dependencies and asymmetric and symmetric layout dependencies of nanostructured matter exhibit agreement between electromagnetic numerical simulations and the theoretical calculations. The explicit representation of optical near-fields by the angular spectrum, which is a decomposition of evanescent waves with different decay lengths, provides a physically intuitive picture in considering the polarization conversion involving subwavelength structures. This study provides insights about unique optical properties, such as unidirectionality, stemming from isotropic shape-engineered nanostructured matter, which will lead to fundamental guiding principles for understanding and engineering nanostructures for novel functionalities.

**ACKNOWLEDGMENTS**

This work was supported in part by the Strategic Information and Communications R&D Promotion Programme (SCOPE) of the Ministry of Internal Affairs and Communications, and by Grants-in-Aid for Scientific Research and Core-to-Core Program, A. Advanced Research Networks from the Japan Society for the Promotion of Science.

# REFERENCES

1. R. J. Potton, "Reciprocity in optics," Rep. Prog. Phys. **67**, 717 (2004).

2. A. Drezet and C. Genet, "Reciprocity and optical chirality," in *Singular and Chiral Nanoplasmonics*, N. Zheludev and S. V. Boriskina, eds. (Pan Stanford Publishing), in press.

3. B. E. A. Saleh and M. C. Teich, *Fundamentals of Photonics*, Chapter 6, (Wiley-Interscience, 2007).

4. H. Dötsch, N. Bahlmann, O. Zhuromskyy, M. Hammer, L. Wilkens, R. Gerhardt, P. Hertel, and A. F. Popkov, "Applications of magneto-optical waveguides in integrated optics: review," J. Opt. Soc. Am. B **22**, 240 (2005).

5. W. Van Parys, B. Moeyersoon, D. Van Thourhout, R. Baets, M. Vanwolleghem, B. Dagens, J. Decobert, O. Le Gouezigou, D. Make, R. Vanheertum, and L. Lagae, "Transverse magnetic mode nonreciprocal propagation in an amplifying AlGaInAs/InP optical waveguide isolator," Appl. Phys. Lett. **88**, 071115 (2006).

6. T. Amemiya, K. Abe, T. Tanemura, T. Mizumoto, Y. Nakano, "Nonreciprocal Polarization Conversion in Asymmetric Magnetooptic Waveguide," IEEE J. Quantum Electron. **46**, 1662 (2010).

7. M. J. Lockyear, A. P. Hibbins, K. R. White, and J. R. Sambles, "One-Way Diffraction Grating," Phys. Rev. E **74**, 056611 (2006).

8. W.-M. Ye, X.-D. Yuan, and C. Zeng, "Unidirectional transmission realized by two nonparallel gratings made of isotropic media," Opt. Lett. **36**, 2842 (2011).

9. J. Xu, C. Cheng, M. Kang, J. Chen, Z. Zheng, Y.-X. Fan, and H.-T. Wang, "Unidirectional optical transmission in dual-metal gratings in the absence of anisotropic and nonlinear materials," Opt. Lett. **36**, 1905 (2011).

10. V. A. Fedotov, P. L. Mladyonov, S. L. Prosvirnin, A. V. Rogacheva, Y. Chen, and N. I. Zheludev, "Asymmetric propagation of electromagnetic waves through a planar chiral structure," Phys. Rev. Lett. **97**, 167401 (2006).

11. A. Drezet, C. Genet, J.-Y. Laluet, and T. W. Ebbesen, "Optical chirality without optical activity: How surface plasmons give a twist to light," Opt. Express **16**, 12559-12570 (2008).

12. C. Menzel, C. Helgert, C. Rockstuhl, E. B. Kley, A. Tünnermann, T. Pertsch, and F. Lederer, "Asymmetric Transmission of Linearly Polarized Light at Optical Metamaterials," Phys. Rev. Lett. **104**, 253902 (2010).

13. C. Cheng, J. Chen, D.-J. Shi, Q.-Y. Wu, F.-F. Ren, J. Xu, Y.-X. Fan, J. Ding, and H.-T. Wang, "Physical Mechanism of Extraordinary Electromagnetic Transmission in Dual-Metallic Grating Structures," Phys. Rev. B **78**, 075406 (2008).

14. E. Wolf and M. Nieto-Vesperinas, "Analyticity of the angular spectrum amplitude of scattered fields and some of its consequences," J. Opt. Soc. Am. A **2**, 886 (1985).

15. T. Inoue and H. Hori, "Representations and Transforms of Vector Field as the Basis of Near-Field Optics," Opt. Rev. **3**, 458-462 (1996).

16. T. Inoue and H. Hori, "Theoretical Treatment of Electric and Magnetic Multipole Radiation Near a Planar Dielectric Surface Based on Angular Spectrum Representation of Vector Field," Opt. Rev. **5**, 295-302 (1998).

17. T. Inoue and H. Hori, "Quantum theory of radiation in optical near field based on quantization of evanescent electromagnetic waves using detector mode," in *Progress in Nano-Electro-Optics IV*, M. Ohtsu ed. (Springer, 2005), pp. 127-199.

18. M. Naruse, T. Inoue, and H. Hori, "Analysis and synthesis of hierarchy in optical near-field interactions at the nanoscale based on angular spectrum," Jpn. J. Appl. Phys. **46**, 6095 (2007).

19. M. Naruse, N. Tate, Y. Ohyagi, M. Hoga, T. Matsumoto, H. Hori, A. Drezet, S. Huant, and M. Ohtsu, "Optical near-field–mediated polarization asymmetry induced by two-layer nanostructures," Opt. Express **21**, 21857 (2013).

20. M. Ohtsu and H. Hori, *Near-Field Nano-Optics* (Kluwer Academic / Plenum Publisher, 1999).

21. D. Jalas, A. Petrov, M. Eich, W. Freude, S. Fan, Z. Yu, R. Baets, M. Popovic, A. Melloni, J. D. Joannopoulos, M. Vanwolleghem, C. R. Doerr, and H. Renner, "What Is – And What Is Not – An Optical Isolator," Nat. Photon. **7**, 579 (2013).

22. R. Carminati, M. Nieto-Vesperinas, and J.-J. Greffet, "Reciprocity of evanescent electromagnetic waves," J. Opt. Soc. A. **15**, 706-712 (1998).

23. N. Tate, H. Sugiyama, M. Naruse, W. Nomura, T. Yatsui, T. Kawazoe, and M. Ohtsu, "Quadrupole–Dipole Transform based on Optical Near-Field Interactions in Engineered Nanostructures," Opt. Express **17**, 11113 (2009).

24. A. Taflove and S. C. Hagness, *Computational Electrodynamics: The Finite-Difference Time-Domain Method* (Artech House, Boston, 2005).

25. D. W. Lynch and W. R. Hunter, "Comments on the Optical Constants of Metals and an Introduction to the Data for Several Metals," in *Handbook of Optical Constants of Solids*, E. D. Palik ed. (Academic Press, Orlando, 1985), pp. 275-367.

**Figure captions**

**Fig. 1.** (a) Geometrical relation of wavevector and polarization vectors. (b) The contour of the integration for the rotation angle α. (c) A schematic illustration of direction-dependent polarization conversion via optical near-field interactions. (d) Schematic diagrams of forward- and backward-dependent polarization conversion based on angular spectrum representation of optical near-fields.

**Fig. 2.** (a) An example of two-layer nanostructures and some associated notations for the geometrical features. (b) A cross-sectional view of the two-layer nanostructures. (c) Polarization conversion efficiency regarding *x*-polarized input to *y*-polarized output light ($T_{X \to Y}^{(F)}$), that regarding *y*-polarized input to *x*-polarized output light ($T_{Y \to X}^{(F)}$) for the forward direction, and those for the backward direction ($T_{X \to Y}^{(B)}$ and $T_{Y \to X}^{(B)}$). (d) The differences in polarization conversion efficiency $\left| T_{X \to Y}^{(F)} - T_{Y \to X}^{(B)} \right|$ and $\left| T_{Y \to X}^{(F)} - T_{X \to Y}^{(B)} \right|$, which correspond to a measure used to quantify unidirectionality.

**Fig. 3.** (a) The average value of $\left| T_{X \to Y}^{(F)} - T_{Y \to X}^{(B)} \right|$, referred to as *average unidirectionality*, in the wavelength region 900–1200 nm as a function of inter-layer gap distance, *G*. (b) The unidirectionality, $\left| T_{X \to Y}^{(F)} - T_{Y \to X}^{(B)} \right|$, with respect to asymmetric and symmetric arrangements of the L-shaped structures in the second layer.

**Fig. 4.** Cross-sectional electric field intensity distributions when the asymmetric and symmetric structures are irradiated with CW input light at a wavelength of 1127 nm.

**Fig. 5.** Theoretical model based on angular spectrum representation. (a) Virtual intermediate layer, which is equidistant from the first and second layers by distance $Z_M$. (b) An array of dipoles induced at the first layer by *x*-polarized input light for the forward direction. (c) An array of dipoles induced at the second layer by *y*-polarized input light for the backward direction. (d) Geometrical illustration for the orientation of dipoles and relative position where the angular spectrum representation is evaluated.

**Fig. 6.** Angular spectrum of (i) the input, (ii) the filter, and (iii) the output with respect to different values of $Z_M$ ($\lambda/20, \lambda/10, \lambda/4$).

**Fig. 7.** The figure-of-merit for the unidirectionality, given by $f^{(F)}_{X_{IN} \to Y_{OUT}} / f^{(B)}_{Y_{IN} \to X_{OUT}}$, as a function of inter-layer distance. An inter-layer distance that is neither too small nor too large provides better unidirectionality, which is consistent with the electromagnetic simulations demonstrated in Fig. 3(a). (b) Angular spectra of the symmetric and asymmetric layouts for the second layer nanostructures. A symmetric layout causes the difference between the angular spectra ($f^{(F)}_{X_{IN} \to Y_{OUT}}$ and $f^{(B)}_{Y_{IN} \to X_{OUT}}$) to diminish, which is consistent with the simulation results shown in Fig. 3(b).

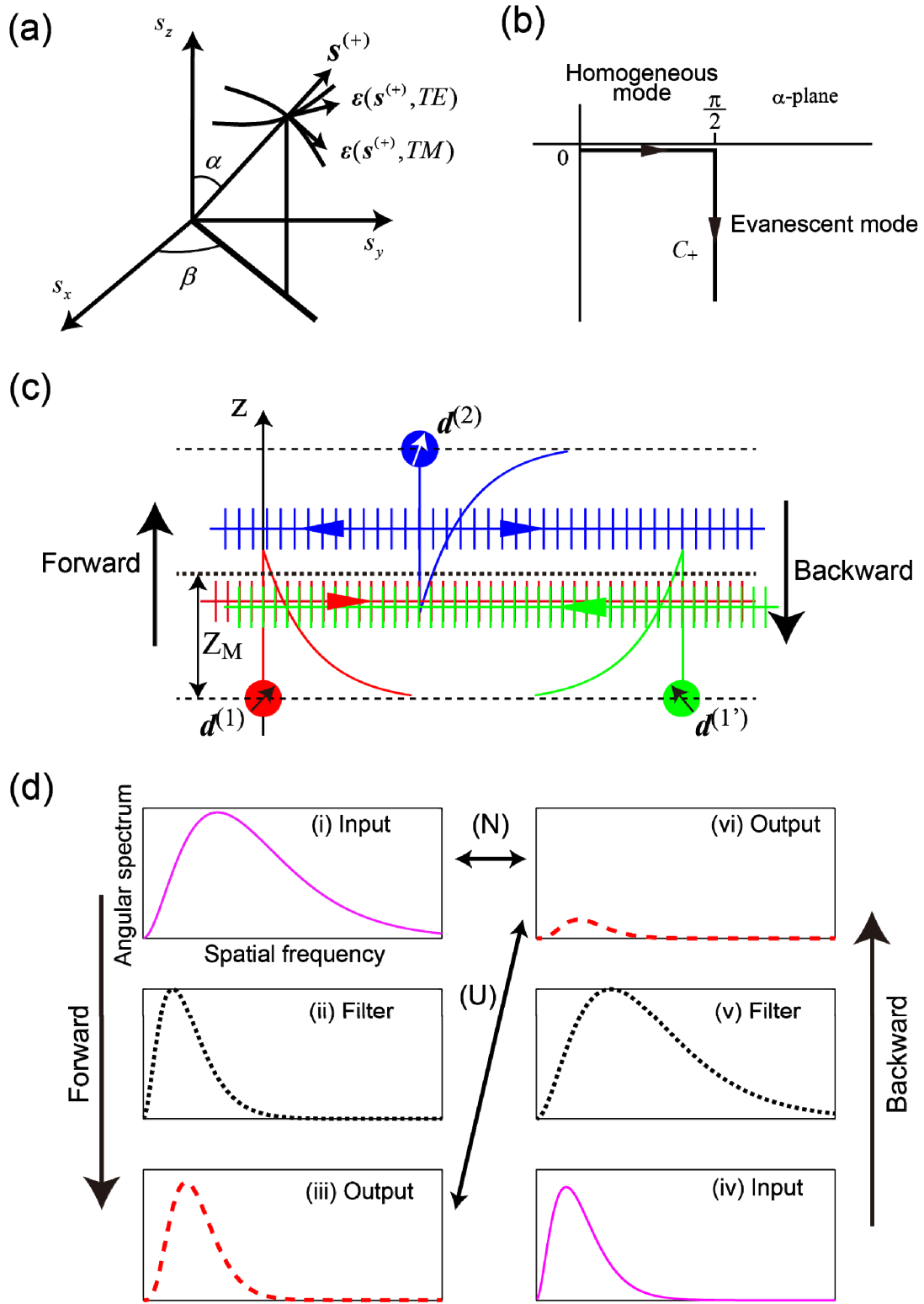

FIG. 1

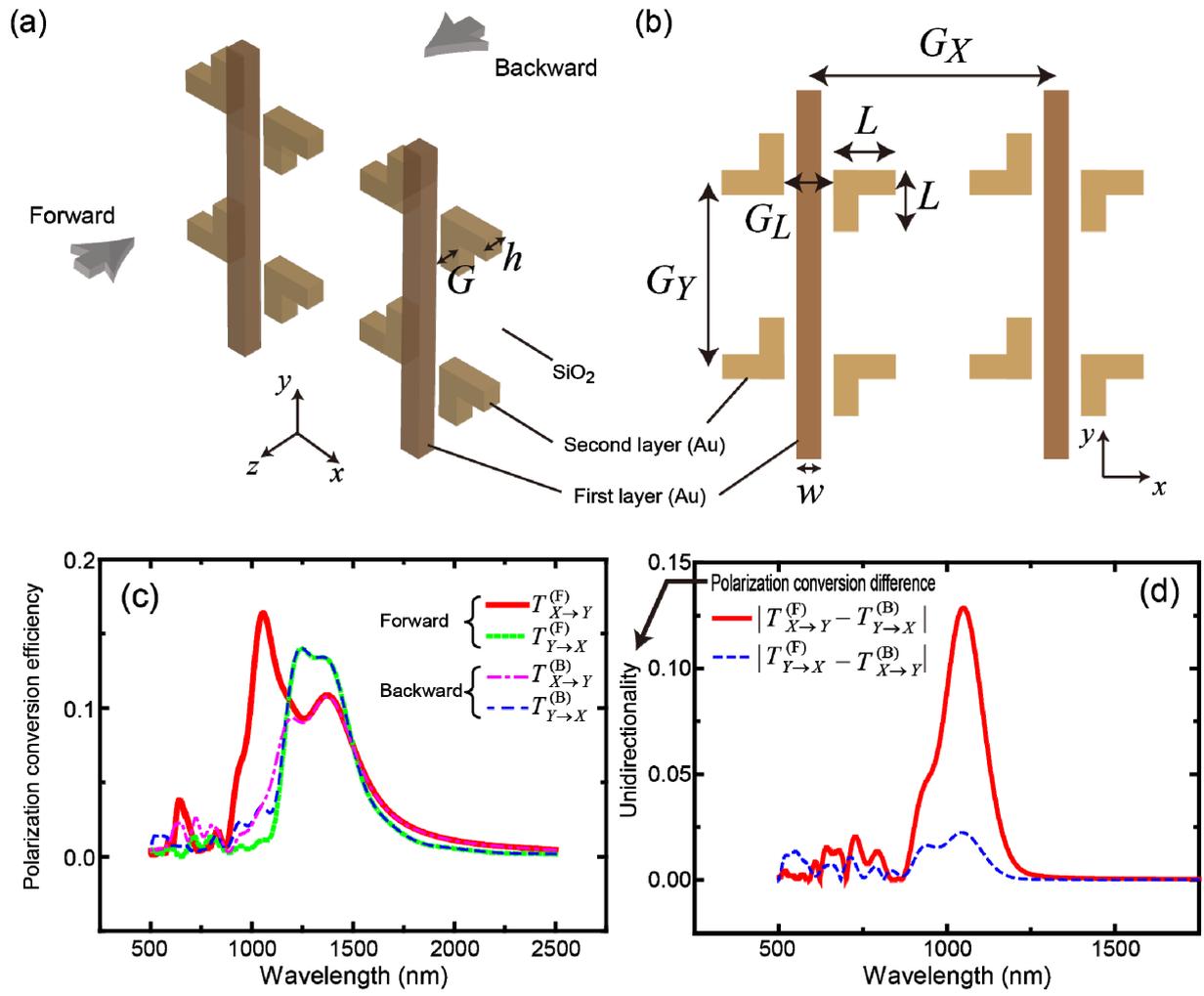

FIG. 2

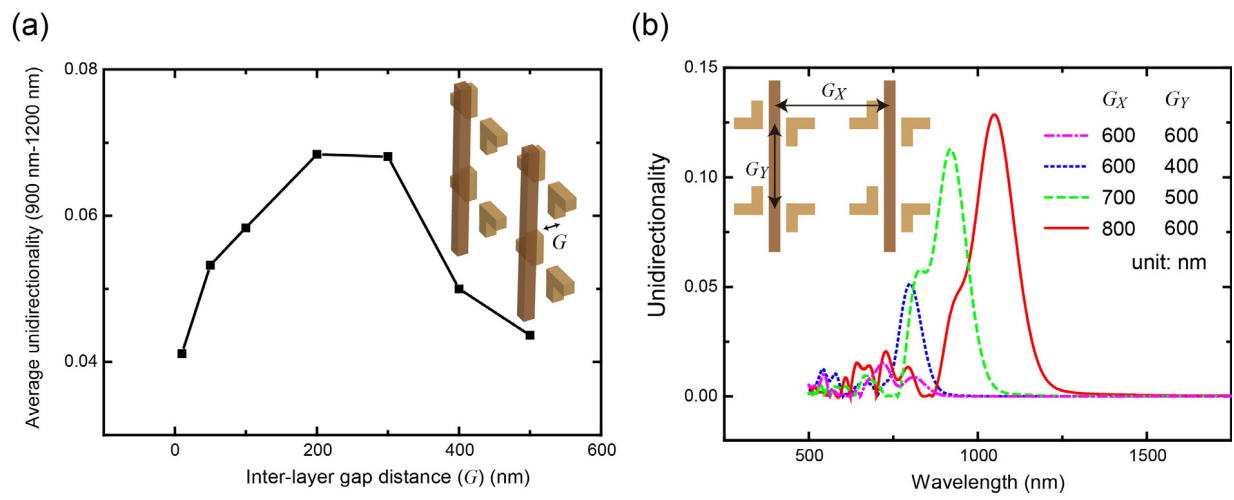

FIG. 3

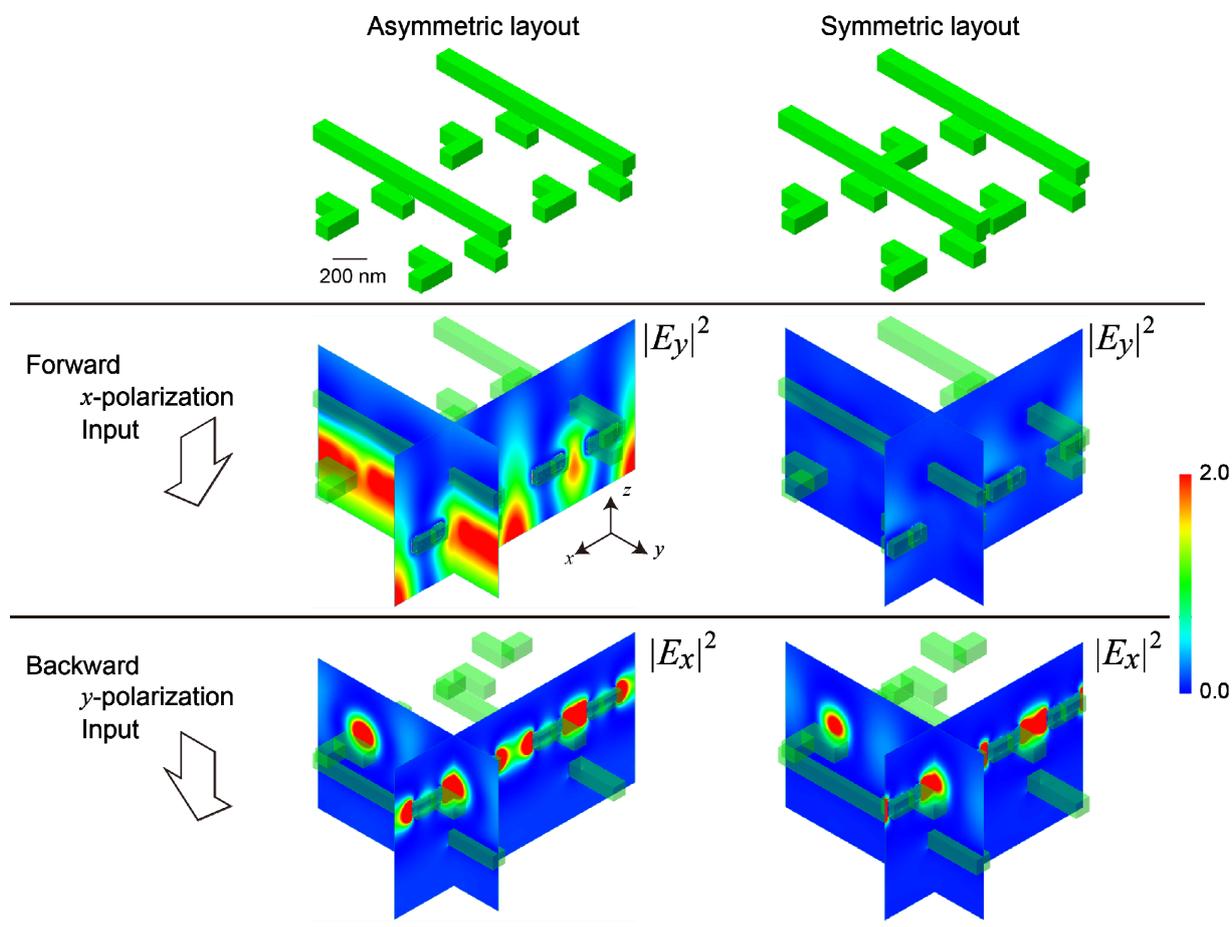

FIG. 4

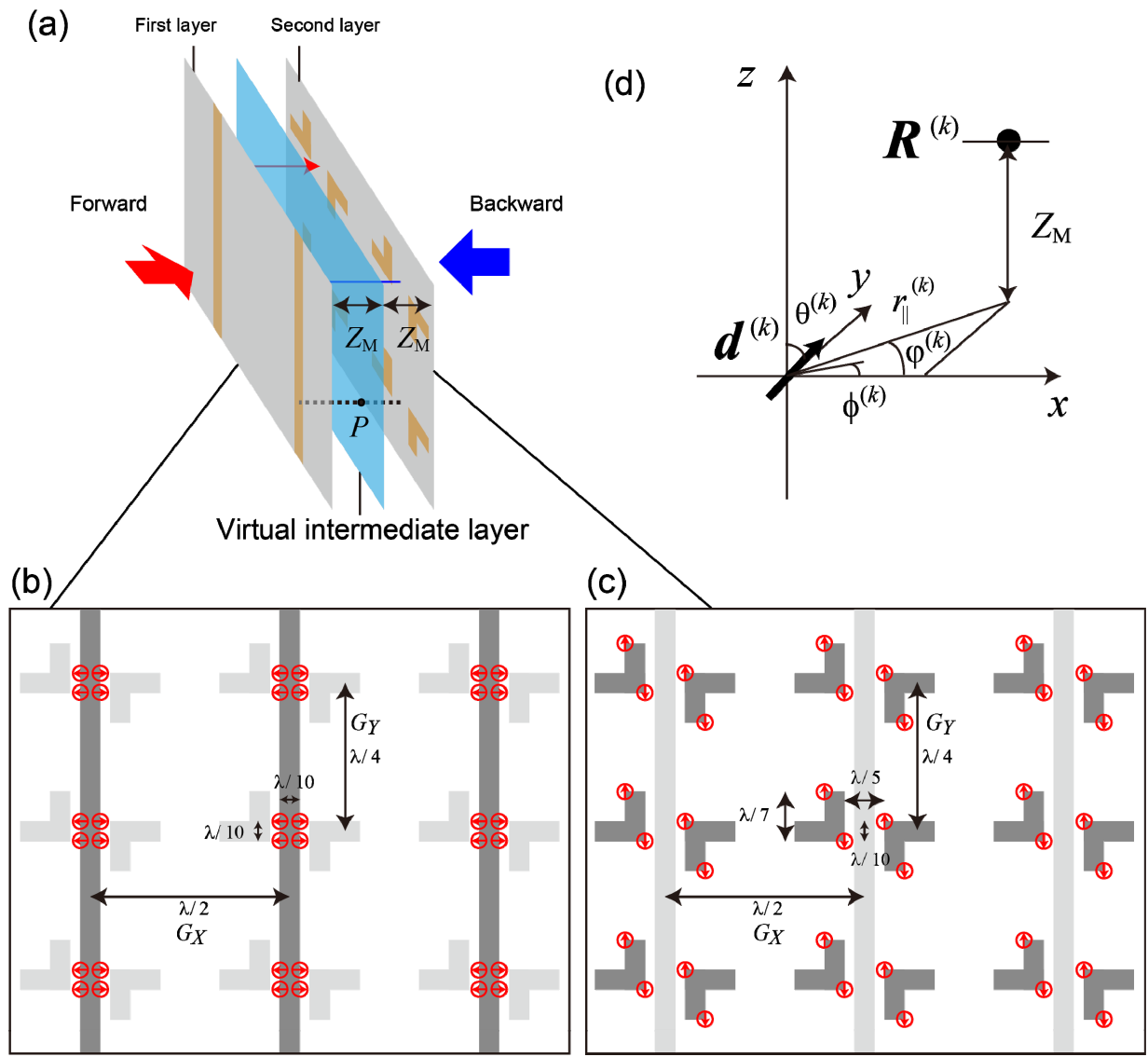

FIG. 5

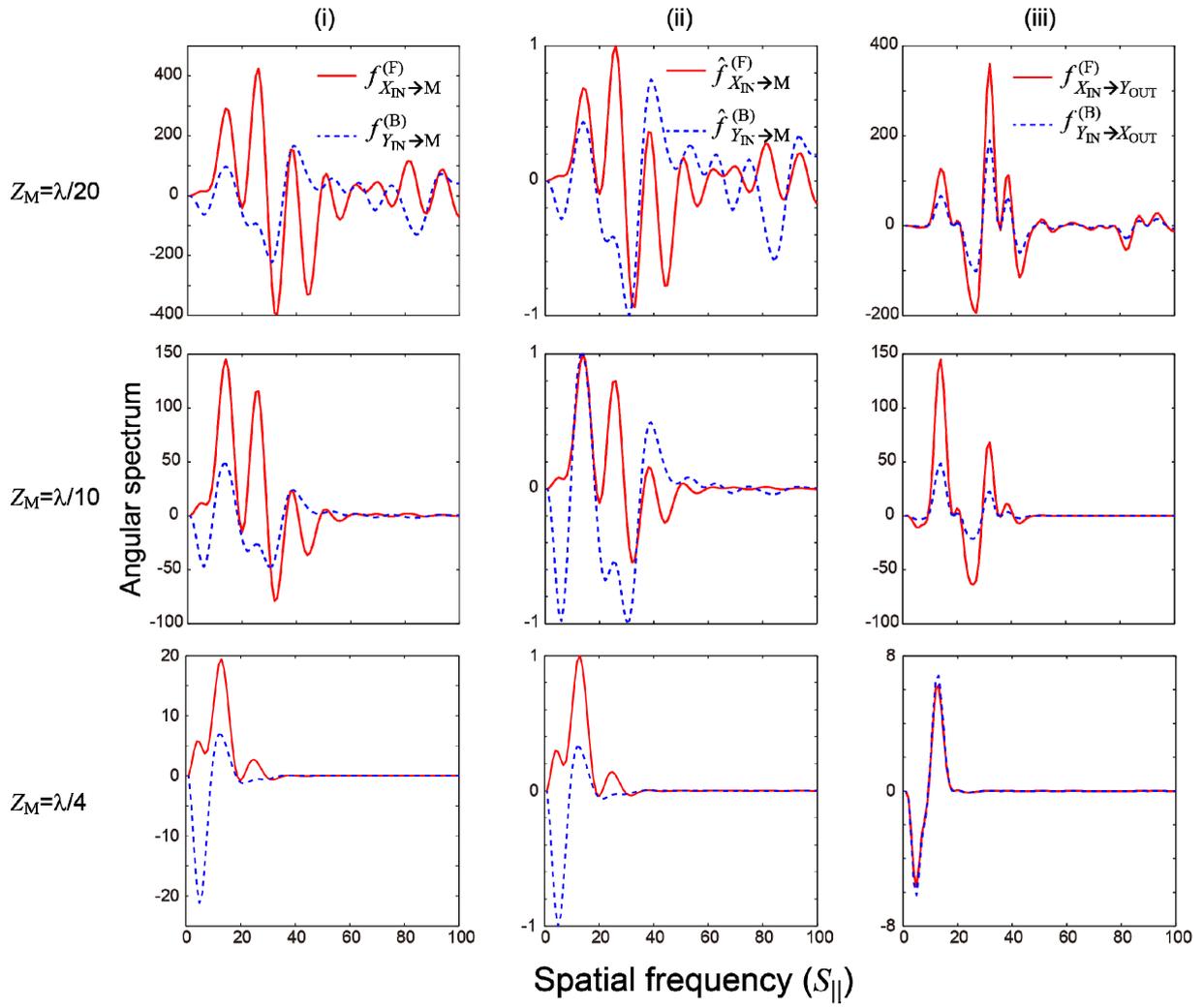

FIG. 6

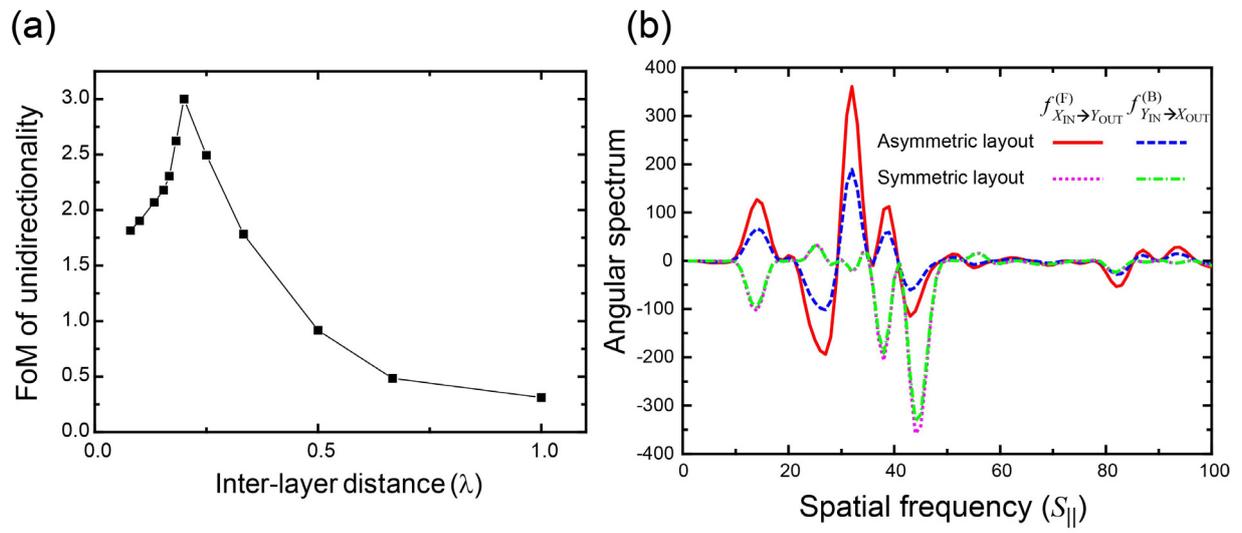

FIG. 7